\documentclass[12pt,a4paper]{iopart}

\usepackage[T1]{fontenc}
\usepackage[latin9]{inputenc}
\usepackage[english]{babel}
\usepackage{color}
\usepackage{float}

\expandafter\let\csname equation*\endcsname=\relax
\expandafter\let\csname endequation*\endcsname=\relax

\usepackage{amsmath}
\usepackage{amssymb,amsthm}

\usepackage{graphicx}
\PassOptionsToPackage{normalem}{ulem}
\usepackage{ulem}
\usepackage{babel}

\numberwithin{equation}{section}
\numberwithin{figure}{section}
\theoremstyle{plain}

  \theoremstyle{plain}
  
  \theoremstyle{definition}
 \newtheorem{defn*}{\protect\definitionname}
 \providecommand{\lemmaname}{Lemma}
  \providecommand{\definitionname}{Definition}
 
\providecommand{\theoremname}{Theorem}

\begin{document}

\title{Writhe induced phase transition in unknotted self-avoiding polygons. }
\author{E Dagrosa$^1$, A L Owczarek$^1$ and T Prellberg$^2$}
\address{$^1$ School of Mathematics and Statistics,
  The University of Melbourne, Parkville, Vic 3010, Australia.}
\address{$^2$ School of Mathematical Sciences, Queen Mary University
  of London, Mile End Road, London E1 4NS, UK.}
\ead{edagrosa@outlook.com, owczarek@unimelb.edu.au, t.prellberg@qmul.ac.uk}

\begin{abstract}
Recently it has been argued that weighting the writhe of unknotted self-avoiding
polygons can be related to possible experiments that turn double stranded DNA. We first solve exactly a directed model and demonstrate that in  such a subset of polygons the problem of weighting their writhe is
associated with a phase transition. We then analyse simulations using
the Wang-Landau algorithm to observe scaling in the fluctuations
of the writhe that is compatible with a second-order phase transition in a undirected self-avoiding polygon model. The transition can be clearly detected when the polygon is stretched with a strong pulling force.
\end{abstract}


\section{Introduction}

Over the past two decades, experiments \cite{Experiment_S007961070000018320000101,edsjsr.288937019960101,edsgcl.1321776019920101,ABRUPT_BUCKLING,edsgcl.29795367720100101,DEUFEL_QUARTZ,1925771620090220}
that turn single molecules of twist-storing polymers like DNA have
been performed. A DNA molecule which is pulled and twisted simultaneously can move from a stretched state into a  state induced by supercoiling \cite{Experiment_S007961070000018320000101}. The conformational transition associated with
these experiments attracts continuing interest. 

\begin{figure}[h]
\centering\includegraphics[scale=0.3]{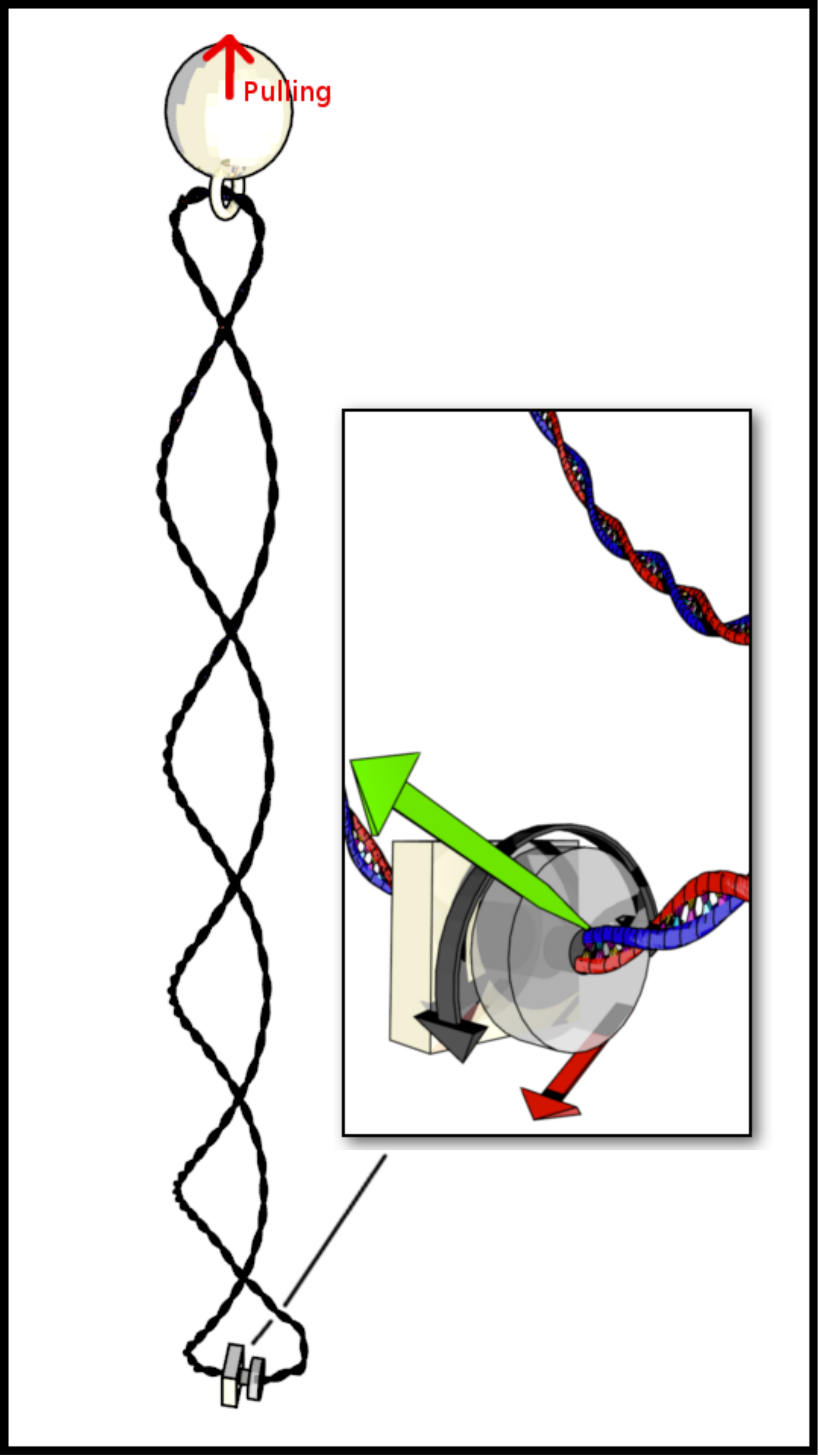}

\caption{\label{fig:TorqueMachine}Sketch of the thought experiment. Both sides
of a DNA like polymer are attached to the apparatus. The depicted
cylinder can be rotated (black arrows). The twist-storing polymer exerts a force $f$
on the cylinder which attacks at a distance $r$ from the centre (green
arrow). On the other hand, a force $F$ (red arrow) can be applied
at a distance $R$ from the centre. Then, the net torque exerted on
the cylinder reads $F\, R-f\, r$.}
\end{figure}

We propose a thought experiment that is slightly different
from the experiments currently performed on DNA. We consider a two sided apparatus
as shown in Figure~\ref{fig:TorqueMachine}. On one side the  twist-storing polymer
is attached to an immobile part of the apparatus, on the other side,
the twist-storing polymer is connected to a turn-able cylindrical structure, which is
used to control the linking number between the two strands of the
double stranded molecule. In order to apply a pulling force, we take
a magnetic bead, attach a loop to it and pass one end of the twist-storing polymer through
the loop before it is attached to the apparatus. Via a magnetic field,
a constant pulling force can be applied to the polymer. Note that
the force does not attack at a specific point of the twist-storing polymer. In fact,
it is important that there is as little as possible interaction between
the loop material and the twist-storing. Instead of controlling the number of
turns, we can imagine to keep the polymer at a constant torque $\beta_{l}$.
This is called the constant torque ensemble.

In the literature \cite{0953-8984-25-46-465102,Fuller_Decomp, Sinha_PhysRevE.85.041802,cond-mat/020349120020324,6010404220110415,2036521120091201,cond-mat/990401819990401}
the mechanics of twist-storing double-stranded DNA has been modeled using ribbons \cite{Fuller_Decomp,1969_WHITE_SL,CALUGAREANU_61_A,CALUGAREANU_61_B}.
The boundary curves of the ribbon can be thought to
represent the strands of a DNA molecule. Only if the knot type
of the polymer is conserved, the linking number between these curves
is controlled by the number of turns and according to the CFW theorem
\cite{Fuller_Decomp,1969_WHITE_SL,CALUGAREANU_61_A,CALUGAREANU_61_B},
the linking number can be absorbed either in the form of twist or
in the form of writhe,
\begin{equation}
Lk=Wr+Tw\label{eq:CFW}\,.
\end{equation}
In this paper, we wish to use a lattice model of ribbons \cite{JvR,Orlandini_TWIST_OF_LatRib}. Conveniently, the linking number of a lattice ribbon was recently proved 
to be equal to the writhe of the centre line of the lattice ribbon in \cite{GENERLARIBBON}, itself a self-avoiding polygon on the sc lattice. In this way, we use a single self-avoiding
polygon instead of the more complicated model of lattice ribbons. While not every 
self-avoiding polygon on the simple cubic lattice can be the centre line of the lattice ribbon, the condition for polygons that are is merely local, so that we 
conjecture that the critical structure remains unaffected by extending the ensemble to all SAPs.
Therefore, we model our thought experiment by weighting the writhe of self-avoiding polygons (SAP) on the simple cubic (sc) lattice.
We are interested in observing the typical signs that are associated with real phase transitions.

\section{Model}

\subsection{The SAW as model of phase transitions}

The self-avoiding walk (SAW) in three dimensions is one of the simplest
models of polymers \cite{Vanderzande}. It reduces polymers down to two features. First,
it represents the fact that a polymer is composed from a large number
of repeating subunits. Second, the self-avoidance reflects the fact
that polymers cannot be compressed arbitrarily. As a model of long
polymers in statistical mechanics, the SAW is known to make correct
predictions about the scaling of the end-to-end distance or equivalently
the radius of gyration of a polymer in a good solvent at thermal equilibrium.
When we denote the length of the SAW by $n$, the expectation value
of the squared radius of gyration is predicted to obey the scaling
form

\begin{equation}
\left\langle R_{g}^{2}\right\rangle _{n}=A_{g}\, n^{2\,\nu}\left(1+b_{g}\, n^{-\Delta}+\ldots\right)\label{eq:R2}
\end{equation}

where $A_{g}$ and $b_{g}$ are constants specific to the lattice
on which the SAW lives. In contrast, the exponents $\nu$ and $\Delta$
are considered to be universal. They have been determined from simulations
\cite{21788719950101} as well as from calculations in the context
of the renormalization group idea. This is due to a mapping of the
SAW onto the magnetic $N$-vector model in the formal limit $N\rightarrow0$.
Recent predictions \cite{clisby} are $\nu=0.587597\pm0.000007$ and
$\Delta=0.528\,\pm0.012$. The predictions for the leading scaling
exponent $\nu$ agree with experiments on DNA \cite{1624176820051007}
that for long molecules of DNA fit their results to $\left\langle R_{g}\right\rangle \sim L^{\nu}$,
where $L$ is the length of the DNA.

In addition SAWs are used as configurations of various standard models of single polymer phase transitions.
Most notable are the collapse and adsorption transitions.
In the latter case, one considers all SAWs above a plane in three
dimensions or above a line in two dimensions and weights the SAWs
by the number of contacts made with the wall (surface or line, respectively).
Denote a SAW by $\varphi=\left\{ \varphi_{i}\right\} _{i=1,..,n+1}$,
where $\varphi_{i}$ are the vertices. Denote by $a$ the number of
visits to the wall, then the SAWs of length $n$ are distributed according
the Boltzmann distribution
\begin{equation}
p\left(\varphi\right)=Z_{n}^{-1}\, e^{\beta_{a}a\left(\varphi\right)}\label{eq:distribution},
\end{equation}
where $a\left(\varphi\right)$ measures the number of visits in $\varphi$,
$Z_{n}=\sum_{\varphi\in\Omega_{n}}e^{\beta_{a}a\left(\varphi\right)}$
is the partition function and $\beta_{a}$ is related to the interaction
strength between the polymer and the wall. We remark that we use the
subscript in the coupling $\beta_{a}$ only as a label, not an index
so that when we denote by $C_{n\, a}$ the number of SAWs of length
$n$ with $a$ visits, we may write the partition function as 
\begin{equation}
Z_{n}\left(\beta_{a}\right)=\sum_{a}C_{n\, a}e^{\beta_{a}\, a}\label{eq:part}.
\end{equation}
The finite size free energy is defined as $f_{n}\left(\beta_{a}\right)=n^{-1}\log Z_{n}\left(\beta_{a}\right)$
so that in the thermodynamic limit $n\rightarrow\infty$, it is known
that there is a critical value $\beta_{a}^{\left(\infty\right)}$
at which the free energy $f=\lim_{n\rightarrow\infty}f_{n}$ becomes
non-analytic. When the critical value is crossed from below the system
undergoes a continuous phase transition from a desorbed phase into
the adsorbed phase. In the desorbed phase the first derivative of
the free energy with respect to $\beta_{a}$ vanishes, while in the
adsorbed phase a certain fraction of the vertices is expected to be
in contact with the wall. Correspondingly, the transition is associated
with a discontinuity in the second derivative $\partial^{2}f\left(\beta_{a}\right)/\partial\beta_{a}^{2}$,
which jumps from zero to some finite value at $\beta_{a}^{\left(\infty\right)}$.
At finite size, the singular part of the free energy near $\beta_{a}^{\left(\infty\right)}$
obeys a standard scaling Ansatz 
\begin{equation}
f_{n}^{\left(sing\right)}\left(\tau\right)=n^{-1}\, h\left(n^{\phi}\tau\right)\label{eq:INTRO_scaling_form}.
\end{equation}
where $\tau=\beta_{a}-\beta_{a}^{\left(\infty\right)}$, $h\left(x\right)$
is a scaling function and $\phi$ is called the crossover exponent.
In the case of the adsorbed walk, the crossover exponent takes a mean
field value of $1/2$ at the transition \cite{DeBell}. An exact solution for certain
directed SAWs in two dimensions is given in \cite{RN09636606120010201},
which also includes an expression for the scaling function. In the
following, we will consider SAP which are SAWs, the first and last
vertex of which are adjacent.

\subsection{The Writhe of SAPs on the sc lattice}

The writhe is a quantity associated with the topology and geometry
of a space curve. For a $C^{1}$ curve
$R$, it is given by \begin{equation}
Wr\left(R\right)=\frac{1}{4\pi}\int_{S^{2}}d\mathbf{d}\, Lk\left(R,\, R+\epsilon\mathbf{d}\right)\label{eq:Wr_as_avg},
\end{equation}
where $Lk\left(R,\,R+\epsilon\mathbf{d}\right)$
is the linking number between the curve $R$
and a copy of $R$ pushed off into the direction of $\mathbf{d}$.

In the lattice polymer literature \cite{Writhe_of_knot,universal_knots,1979122520060214,4472730720091021,cond-mat/070378420070329},
the writhe of a SAP is usually computed based on a theorem by Laing and Sumners \cite{SUMNERS_WRITHES},
who realized that the expression for the writhe simplifies for a SAP $\varphi$ on the sc lattice by considering the
linking number between $\varphi$ and four copies of it translated by $\mathbf{d}=\left(\pm1/2,\,\pm1/2,\,\pm1/2\right)$, leading to
 \begin{equation}
Wr\left(\varphi\right)=\frac{1}{4}\,\left\{ Lk\left(\varphi,\,\varphi+\frac{1}{2}\left(1,\,1,\,1\right)^{T}\right)+Lk\left(\varphi,\,\varphi+\frac{1}{2}\left(1,\,-1,\,1\right)^{T}\right)+...\right\} .\label{eq:Wr_phi}
\end{equation}
It was shown in \cite{DagrosaThesis} that this can further be rewritten as 
\begin{equation}
Wr\left(\varphi\right) = Lk\left(\varphi,\,\varphi+\frac{1}{2}(1,1,1)^T\right)-\frac{1}{8}\sum_{i=1}^{n}\left[\hat{\varphi}_{i-1},\,\hat{\varphi}_{i},\,(1,1,1)^T\right]\left\langle (1,1,1)^T,\,\hat{\varphi}_{i-1}+\hat{\varphi}_{i}\right\rangle \label{eq:WR_sc_writhe_twist_formula}.
\end{equation}
where the square brackets denote the triple product and $\hat{\varphi_i}$ denotes the unit vector from $\varphi_i$ to $\varphi_{i+1}$. It is this latter formula that will be used in our simulations.
\subsection{Model: Self-Avoiding Unknot on the sc lattice}

\begin{figure}
\centering\includegraphics[width=5cm]{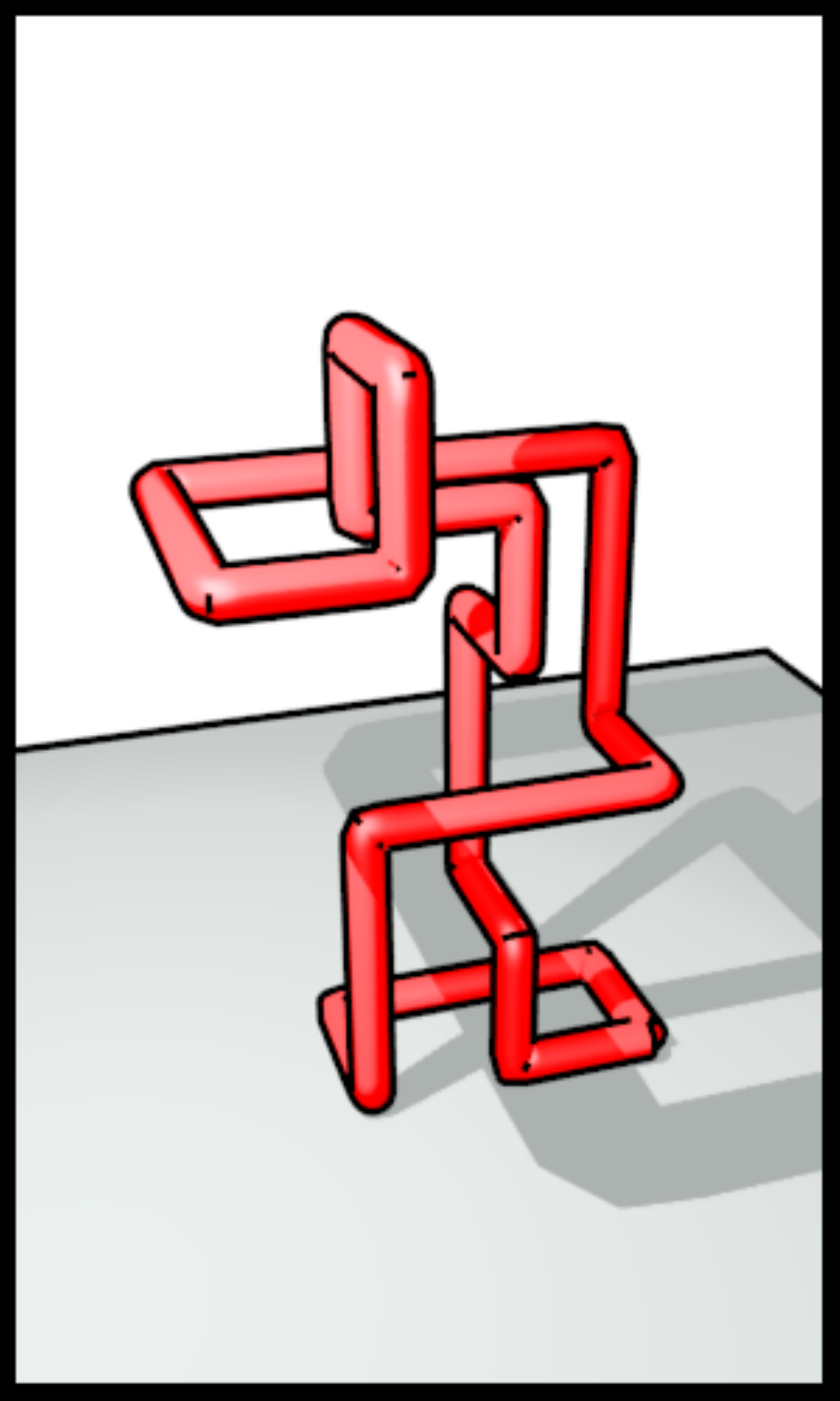}\caption{\label{fig:SAUK_SAP_picture}A self-avoiding unknot (SAUK) above a
surface of length $n=28$. The step furthest away from the surface
defines the extension $h$. }
\end{figure}

As mentioned in the introduction, we consider a model in which we weight the writhe of self-avoiding
polygons on the sc lattice. To model our thought experiment, we need to restrict the knot type of the self-avoiding polygons
to be the unknot. We refer to these as self-avoiding unknots
(SAUK). Imagine a surface at $x_{3}=0$, so that the SAUK is tethered to the origin and restricted
to the positive half space $\left(\varphi_{i}\right)_{3}\geq0$. In
the thought experiment, the energy associated with pulling is proportional
to the distance $h$ of the bead from the plane $x_{3}=0$. At least
for strong pulling forces this should typically be identical to the
$x_{3}$-component of the vertex furthest away from the surface. We
define
\begin{equation}
h=\max_{i=1,..,n}\left(\varphi_{i}\right)_{3}.\label{eq:WL_h-1}
\end{equation}
The second parameter $w$ is an appropriate multiple of the writhe.
Then, the partition function reads
\begin{equation}
Z_{n}\left(\beta_{l},\,\beta_{h}\right)=\sum_{w,\, h}C_{n\, w\, h}e^{\beta_{l}w+\beta_{h}h}\label{eq:SAUK_Z-1}\,,
\end{equation}
where $C_{n\, w\, h}$ is the number of SAUKs of length $n$, writhe
$w$ and extension $h$. The finite size free energy is given by $f_{n}\left(\beta_{l},\,\beta_{h}\right)=n^{-1}\log\, Z_{n}\left(\beta_{l},\,\beta_{h}\right)$.

\subsection{Overview}

The remainder of this paper is structured as follows.
In Section~\ref{sec:Directed-SAUK-with_AXIS} we will define a directed
version of a SAUK. We use it to show via an exact solution that weighting
the writhe of certain unknotted polygons induces a real phase transition. By
construction, this transition must be associated with the SAUK wrapping
around itself. In Section~3 we use simulations to weight the writhe
of pulled SAUK on the sc lattice. We summarize and discuss the results
in Section~4.


\section{An exactly solved model of twist-storing polymers\label{sec:Directed-SAUK-with_AXIS}}

In this section, we solve a directed version of a SAUK on the sc lattice, weighted by its writhe.
When a general SAUK is under strong stretching force and not too much torque, we might
expect that a typical conformation may be approximated by two SAWs,
directed in the pulling direction, that begin at the origin and join at a point on
some plane $x_{3}=h$. The directedness of the SAWs guarantees that
the joined object is indeed an unknot. Hence, to model the strong pulling regime it may be sensible to consider twist-storing polymers 
using these configurations. However, on any fixed $x_3$-plane one still has two non-intersecting self-avoiding walks
to consider. Not surprisingly, such a model is resistent to being solved directly. We simplify the model in two further steps,
firstly restricting one of the SAWs to be a simple straight line along the pulling direction $\hat{x}_{3}$, and secondly restricting the other SAW by
confining it to a thin slab and further restricting the allowed steps. The thus simplified model turns out to be exactly solvable and is shown 
to have a phase transition on varying the torque fugacity.

\begin{figure}
\centering\includegraphics[width=10cm,height=6cm]{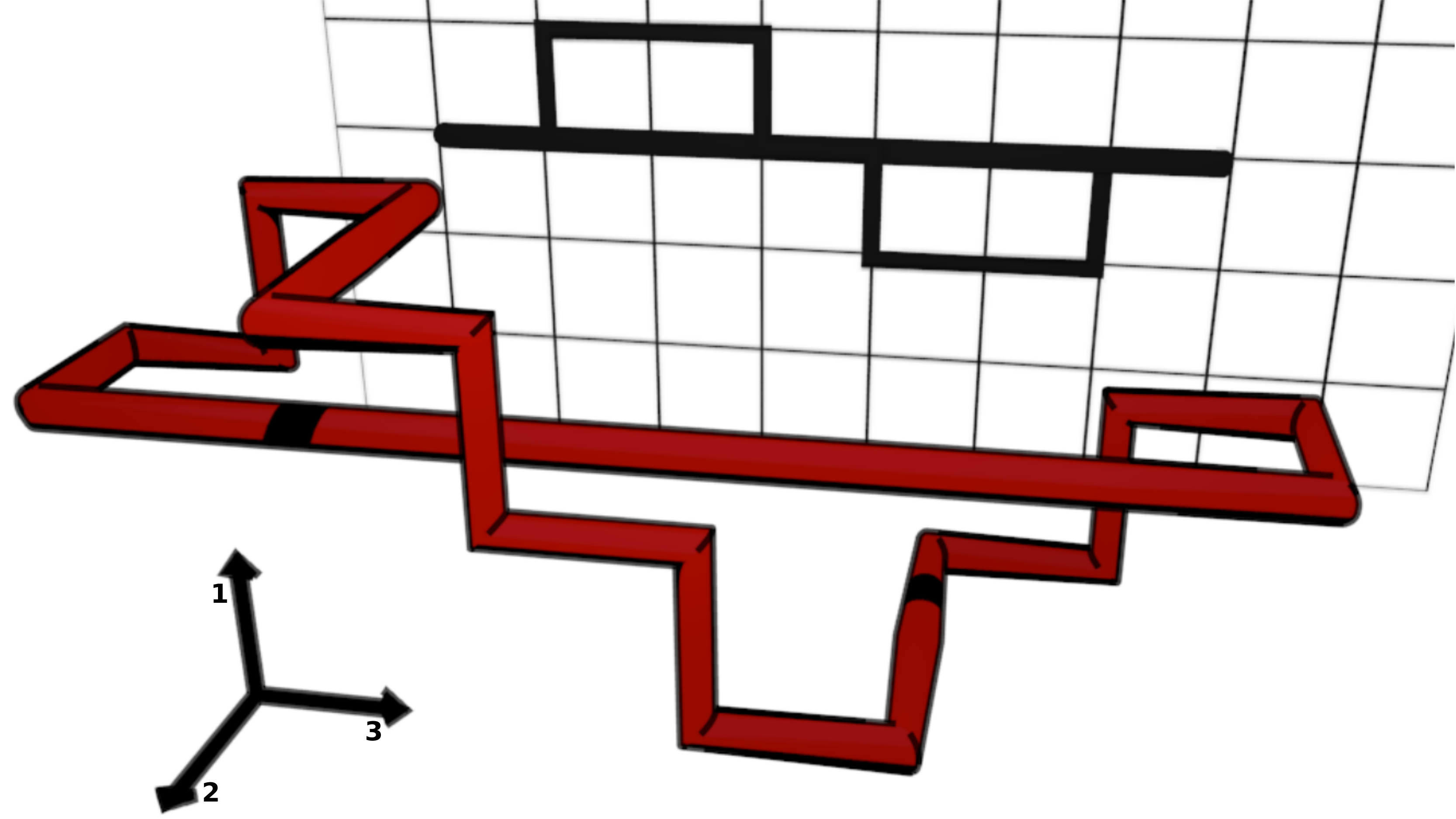}

\caption{\label{fig:WW_SAP-with-axis}Directed SAUK with axis of length 24.}
\end{figure}

When restricting one of the SAWs to step only along the pulling direction axis, the resulting object shall be called a directed SAUK with axis. 
An example of such a directed SAUK with axis (which is also constrained according to the rules defined below) is shown in
Figure~\ref{fig:WW_SAP-with-axis}. More precisely, choosing the lattice base vector $\hat{x}_{3}$
on the sc lattice and a root vertex $\varphi_{1}$ , we define a directed
SAUK with axis $\varphi$ of length $n$ as a two component lattice
object. The first component is a directed self-avoiding walk comprising
of vertices $\varphi_{1},\,\varphi_{2},\,...\varphi_{m}$ such that 
\begin{enumerate}
\item The walk ends on the axis defined by $\hat{x}_{3}$ and the root,
i.e. $\varphi_{m}=\varphi_{1}+\left(n-m+1\right)\,\hat{x}_{3}$,
\item No vertex of the walk except $\varphi_{1}$ and $\varphi_{m}$ lie
on the axis, i.e. $\varphi_{i}\neq\,\left(\varphi_{1}+j\,\hat{x}_{3}\right)$
for all $j\in\mathbb{N}$ and all $i=2,..,m-1$,
\item The walk is directed and makes no steps against the $\hat{x}_{3}$
direction, i.e. $\left(\varphi_{i+1}-\varphi_{i}\right)\cdot\hat{x}_{3}\geq0$
for all $i=2,..,m-1$.
\end{enumerate}
The second component is a walk of length $n-m$ that connects the
vertices $\varphi_{m}$ and $\varphi_{1}$. The vertex positions are
given as $\varphi_{i}=\varphi_{1}+\left(n-i+1\right)\,\hat{x}_{3}$ for all
$i=m+1,...,n-1$. 

We now construct a restricted directed SAUK with axis as follows. 
We consider a walk starting at the origin with initial steps $(-\hat x_2)\hat x_3$.
We then grow this walk by
choosing to append steps to the final vertex as follows. We can append double steps $\hat{x}_{1}\hat{x}_{3}$ or $\left(-\hat{x}_{1}\right)\hat{x}_{3}$. 
Additionally, if the $x_1$-component of the final vertex is non-zero, we can also to append triple steps $(-\hat{x}_{2})(-\hat{x}_{2})\hat{x}_{3}$ or $\hat{x}_{2}\hat{x}_{2}\hat{x}_{3}$ if the $x_2$-component of the final vertex is equal to $1$ or $-1$, respectively. 
A walk constructed in such a way lies in the slab $-1\leq x_2\leq 1$. With
every appended double or triple step, we also prepend a step $-\hat{x}_{3}$
to the starting vertex of the walk. Whenever a vertex with components $\left(0,\,-1,\, n_{3}\in\mathbb{Z}\right)$
is reached, we can generate a SAUK by appending a double step in $\hat{x}_{2}(-\hat x_3)$.
A lattice object constructed according to these rules is shown in
Figure~\ref{fig:WW_SAP-with-axis}. 

Considering the writhe formula (\ref{eq:WR_sc_writhe_twist_formula}) we see that when $x_{1}>0$ the
double steps in $\hat{x}_{2}$-direction increase the writhe
$w$ by $1$, and the double steps in $-\hat{x}_{2}$-direction
decrease the writhe by $1$. When $x_{1}<0$, the writhe contributions
are inverted. There is no other source of writhe. 
Note that the writhe of this simplified directed SAUK with axis is an even integer. We do not consider a pulling force, so that when
we denote the number of these specified SAUKs with length $n$ and
writhe $w$ by $C_{n\, w}$ the partition function reads 
\begin{equation}
Z_{n}\left(\beta_{l}\right)  =  \sum_{w}C_{n\, w}e^{\beta_{l}w}\label{eq:WRP_Zn}.
\end{equation}
We note that we here use $w=Wr$, in contrast to the simulations of the undirected SAUK, where it is advantageous to use $w=4Wr$.

To solve the model, we will obtain an expression for the generating
function of restricted SAUKs with axis
\begin{equation}
\bar{a}\left(\beta_{l},\, z\right):=\sum_{n}Z_{n}\left(\beta_{l}\right)z^{n}\label{eq:WW_GEN_SAUK}
\end{equation}
and analyse its singularities, since the closest singularity $z_c(\beta_l)$ to the origin gives us the free energy as $f(\beta_l)=-\log z_c(\beta_l)$. 
The generating function involved here is an algebraic function and as such can only have poles and branch points as singularities. We will find that
for one range of $\beta_l$ the closest singularity to the origin is given by a branch point, while for another region it is given by a simple pole. The changeover
between these two regions gives a phase transition. In particular, for small $\beta_l$ where we expect a phase of low writhe, we find a branch point singularity
and a temperature-independent free energy, which in turn implies $\langle w\rangle=o(n)$. On the other hand, for large $\beta_l$ we find a simple pole and a 
non-constant free energy, which implies $\langle w\rangle\sim C n$ with temperature-dependent $C>0$.

The strategies for solving similar problems have been used for example
in \cite{1742-5468-2005-05-P05008,OWCZAREK_3D_STICKY_WALL}. In addition
to the generating function~\ref{eq:WW_GEN_SAUK}, one introduces
a generating function for unfinished SAUKs of length $n$, where
$n$ counts the number of steps in the walk and in the axis. Denote
by $a_{h}$ the generating function of unfinished SAUKs whose walk
ends on the plane $x_{2}=-1$ at $x_{1}=h$. Correspondingly, when
the walk ends on the plane $x_{2}=1$, we denote the generating function
by $b_{h}$. Therefore, $\bar{a}\left(\beta_{l},\, z\right)=z\, a_{0}\left(\beta_{l},\, z\right)$.
Define $y_{l}=\exp\left\{ \beta_{l}\right\} $, and with 
\[
\Theta\left(x\right)=\begin{cases}
1 & x>0,\\
0 & x\leq0,
\end{cases}
\]
define
\begin{equation}
H{}_{h}\left(y_{l}\right):=\left\{ y_{l}^{-1}\,\Theta\left(h\right)+y_{l}\,\Theta\left(-h\right)\right\} \label{eq:WW_Hh}.
\end{equation}
Note that $H_{-h}=1/H_{h}$ (when $h\neq0$) and $H_{0}\left(y_{l}\right)=0$.
Then, the recursion relations for the generating functions read 
\begin{equation}
a_{h}=z^{3}\left(\delta_{h,0}+a_{h+1}+a_{h-1}\right)+z^{4}\, H_{h}\, b_{h}\label{eq:WW_ah},
\end{equation}
and
\begin{equation}
b_{h}=z^{3}\left(b_{h+1}+b_{h-1}\right)+z^{4}\, H_{-h}\, a_{h}\label{eq:WW_bh}.
\end{equation}
The factors $z^{3}$ and $z^{4}$ correspond to double and triple
steps, respectively. The additional factor of $z$ is associated with
the additional reverse step on the axis. For $h=0$, the generating
function $a_{0}$ also contains the initial configuration formed by
three steps. The goal is to obtain $a_{0}$ and to get rid of all
the $b_{h}$. Note that for $h\neq0$ (\ref{eq:WW_ah}) can be inverted
to give
\begin{equation}
b_{h}=z^{-4}\, H_{-h}\left\{ a_{h}-z^{3}\left(a_{h+1}+a_{h-1}\right)\right\} \label{eq:WW_bh_inv}.
\end{equation}
For $\left|h\right|>1$, this can be used in (\ref{eq:WW_bh}) to
obtain the bulk relation for $a_{h}$,
\begin{equation}
0=z^{6}a_{h-2}-2z^{3}a_{h-1}+\left(1+2\, z^{6}-z^{8}\right)a_{h}-2\, z^{3}a_{h+1}+z^{6}\, a_{h+2}\label{eq:WW_diff}.
\end{equation}
This is a difference equation of order four for $a_{h}$. The general
solution has the form $a_{h}=\sum_{i=1}^{4}C_{i}\lambda_{i}^{h}$,
where $\lambda_{i}$ are the roots of the characteristic equation
to~(\ref{eq:WW_diff}). The series expansion of the solutions yields
$\lambda_{1}=1/\lambda_{2}=z^{3}+z^{7}+z^{9}+O\left(z^{11}\right)$
and $\lambda_{3}=1/\lambda_{4}=z^{3}-z^{7}+z^{9}+O\left(z^{11}\right)$.
The solutions with the alternating sign in the expansion are incompatible
with a generating function (i.e. $Z_{n}>0$) so that the solution
must take the form
\begin{equation}
a_{h}=C^{+}\lambda_{+}^{h}\Theta\left(h\right)+C^{-}\lambda_{-}^{h}\Theta\left(-h\right)\label{eq:WW_ah_2}
\end{equation}
with 
\begin{equation}
\lambda_{\pm}\left(z\right)=\frac{1}{2z^{3}}\left(1-z^{4}\mp R\left(z\right)\right),
\end{equation}
where $R\left(z\right)=\sqrt{1-2z^{4}-4z^{6}+z^{8}}$ and $\lambda_{+}\lambda_{-}=1$.
It follows from the series expansions that $\lambda_{+}$ applies
for $h>0$ and $\lambda_{-}$ for $h<0$. 

With $r=\sqrt{\frac{1}{3}\left\{ 3-\frac{2\ 6^{2/3}}{\left(-9+\sqrt{129}\right)^{1/3}}+\left(6\left(-9+\sqrt{129}\right)\right)^{1/3}\right\} }$
, the solution~(\ref{eq:WW_ah_2}) has a square root singularity
at 
\begin{eqnarray}
z_{b} & = & \frac{1}{2}\left(\sqrt{3+\frac{2}{r}-r^{2}}-1-r\right)\label{eq:zb}\\
 & \approx & 0.717\nonumber .
\end{eqnarray}
Therefore $R\left(z_{b}\right)=0$ and $z_{b}$ can be associated
with a low-writhe phase as described above. For $\left|h\right|>1$ the solution (\ref{eq:WW_ah_2})
holds for any $C^{+},\, C^{-}\neq0$. To make the solution work at
$\left|h\right|=1$, one requires two boundary conditions that include
$a_{1}$ and $a_{-1}$. These are obtained by considering clock- and
counterclockwise round trips around the axis. According to (\ref{eq:WW_ah})
\begin{equation}
a_{-1}=a_{-1}\left(a_{-2},\, a_{0},\, b_{-1}\right)\label{eq:WW_a1}.
\end{equation}
This notation implies that $a_{-1}$ is expressed through $a_{-2},\, a_{0}$
and $\, b_{-1}$. One continues using the relations (\ref{eq:WW_ah}),
(\ref{eq:WW_bh}) and (\ref{eq:WW_bh_inv}) on the right-hand side
of~(\ref{eq:WW_a1}). For example, in the next step, use (\ref{eq:WW_bh})
to express $b_{-1}=b_{-1}\left(b_{0},\, b_{-2},\, a_{-1}\right)$.
It follows an equation 
\begin{equation}
a_{-1}=a_{-1}\left(a_{-2},\, a_{0},\, b_{0},\, b_{-2},a_{-1}\right).
\end{equation}
Use (\ref{eq:WW_bh}) again to substitute for $b_{0}$ yielding $a_{-1}=a_{-1}\left(a_{-2},\, a_{0},\, b_{-1},\, b_{1},\, b_{-2},a_{-1}\right)$.
Next, all $b_{h}$ can be expressed in terms of $a_{h}$. It follows
\newline $a_{-1}=a_{-1}\left(a_{-2},\, a_{0},\, a_{-2},a_{-1},a_{1},a_{2},\, a_{-3},a_{-1}\right)$.
Finally, with (\ref{eq:WW_ah}), $a_{0}$ is expressed through $a_{-1}$
and $a_{1}$ yielding the boundary condition
\begin{eqnarray}
0 & = & z^{6}-z^{12}\left(1+y_{l}^{2}\right)-z^{6}a_{-3}-z^{3}\left(-2+z^{6}\right)a_{-2}\nonumber \\
 &  & +\left(-1+z^{6}+z^{8}-z^{12}\left(1+y_{l}^{2}\right)\right)a_{-1}\nonumber \\
 &  & -z^{6}\left(-1+z^{6}\right)\left(1+y_{l}^{2}\right)a_{1}-z^{9}y_{l}^{2}a_{2}\label{eq:WW_bnd1}.
\end{eqnarray}
The second boundary condition is obtained analogously by starting
with $a_{1}=a_{1}\left(a_{2},\, a_{0},\, b_{1}\right)$. It reads
\begin{eqnarray}
0 & = & z^{12}\left(1+y_{l}^{2}\right)-z^{6}y_{l}^{2}+z^{9}a_{-2}+z^{6}\left(-1+z^{6}\right)\left(1+y_{l}^{2}\right)a_{-1}\nonumber \\
 &  & +\left(z^{12}+\left(1-z^{6}-z^{8}+z^{12}\right)y_{l}^{2}\right)a_{1}\nonumber \\
 & & +z^{3}\left(-2+z^{6}\right)y_{l}^{2}a_{2}+z^{6}y_{l}^{2}a_{3}\label{eq:WW_bnd2}.
\end{eqnarray}
When the Ansatz (\ref{eq:WW_ah_2}) is plugged into the boundary relations
(\ref{eq:WW_bnd1}, \ref{eq:WW_bnd2}), one may solve the equation
system for $C^{\pm}$. With the expression for $a_{0}$ from (\ref{eq:WW_ah})
the generating function becomes
\begin{eqnarray}
\bar{a}\left(z,\, y_{l}\right) & = & z\, a_{0}\\
 & = & z^{4}\left(1+C^{-}\lambda_{-}^{-1}+C^{+}\lambda_{+}\right).
\end{eqnarray}
The denominator $D$ of $\bar{a}$ can be considered a function $D=D\left(\lambda_{+}\left(z\right),\,\lambda_{-}\left(z\right),\, z,\, y_{l}\right)$
the roots of which are the pole singularities of $\bar{a}$. However,
$D\left(z\right)$ is a very complicated function of $z$, so it seems
not possible to determine the roots of $D$ and thereby the free energy
in the corresponding phases. Nevertheless, one may set $z=z_{b}$
and determine if there exist values $y_{l}^{\left(c\right)}$, so
that $D\left(z=z_{b},\, y_{l}^{\left(c\right)}\right)=0$. This means
that there exists a phase transition from the low-writhe phase into a phase
associated with a pole singularity, which can be seen as a high-writhe phase. At $z=z_{b}$, $\lambda_{+}=\lambda_{-}=\left(1-z_{b}^{4}\right)/z_{b}^{3}$
and as a function of $y_{l}$, the denominator has the form
\begin{equation}
D\left(z=z_{b},\, y_{l}\right)=A_{0}\left(z_{b}\right)-A_{2}\left(z_{b}\right)y_{l}^{2}+A_{4}\left(z_{b}\right)y_{l}^{4},
\end{equation}
where 
\begin{equation}
A_{0}=4z_{b}^{18}\left(1+z_{b}^{2}-6z_{b}^{6}-z_{b}^{8}+z_{b}^{10}\right),
\end{equation}
\begin{eqnarray}
A_{2} & = & \left(1+z_{b}^{4}\right)\times\nonumber \\
 &  & \left\{ 5-z_{b}^{4}\left(17+41z_{b}^{2}-18z_{b}^{4}-93z_{b}^{6}-87z_{b}^{8}+67z_{b}^{10}\right.\right.\nonumber \\
 &  & \left.\left.+100z_{b}^{12}+z_{b}^{14}-54z_{b}^{16}-8z_{b}^{18}+7z_{b}^{20}\right)\right\} ,
\end{eqnarray}
and
\begin{equation}
A_{4}=z_{b}^{12}\left(-1+z_{b}^{2}\right){}^{4}\left(1+z_{b}^{2}\right){}^{2}\left(1+2z_{b}^{2}+3z_{b}^{4}\right).
\end{equation}

The singularities closest to the origin are found when
\begin{equation}
y_{l}^{\pm}=\sqrt{\frac{A_{2}\pm\sqrt{A_{2}^{2}-4A_{0}A_{4}}}{2\, A_{4}}}.
\end{equation}
These two solutions are reciprocal of each other, and hence $\beta_{l}^{\pm}=\log y_{l}^{\pm}$ differ only by a change of sign.
We denote by $z_{c}^{\pm}\left(\beta_{l}\right)$
the pole singularities as a function of $\beta_{l}$ for which $z^{\pm}\left(\beta_{l}^{\pm}\right)=z_{b}$.
At $\beta_{l}=0$ the system is in the low-writhe phase so that the free
energy must take the form 
\begin{equation}
f\left(\beta_{l}\right)=-\begin{cases}
\log z_{c}^{-}\left(\beta_{l}\right) & \beta_{l}<\beta_{l}^{-},\\
\log\left(z_{b}\right) & \mbox{\ensuremath{\beta}}_{l}^{-}\leq\beta_{l}\leq\mbox{\ensuremath{\beta}}_{l}^{+},\\
\log z_{c}^{+}\left(\beta_{l}\right) & \beta_{l}>\beta_{l}^{+}.
\end{cases}\label{eq:WW_free_energy}
\end{equation}
Numerically, the values approximate to ($z_{b}=0.7167$) 
\begin{eqnarray}
\beta_{l}^{\pm} & = & \pm1.4045\label{eq:WW_betaC}.
\end{eqnarray}

We conclude that we have shown that weighting the writhe
of restricted directed SAUK with axis induces a phase transition at a non-trivial value of $\beta_l$. 
By construction, the high torque phase is associated with SAUKs that wrap around their
axis. By analysing the singularity of the free energy one can further show that the phase transition is a second-order transition
with a jump-discontinuity in the second derivative.

\section{Simulations}

Self-avoiding lattice knots, including SAUKs have been treated
via simulations in the lattice polymer literature before. Most notably
in \cite{universal_knots}, the authors considered SAPs of lengths
up to $2\times10^{5}$ via Markov Chain Monte Carlo (MCMC) simulations,
where (effectively) uncorrelated samples were generated with the two-point
pivot algorithm. Via a knot detection algorithm, the samples could
then be categorized by knot type. Their main result was that the scaling
exponent associated with the radius of gyration appears to be invariant
when the ensemble of SAP is restricted to a certain knot type. Such a result
cannot be derived from the renormalization group, as the renormalization
group transformation will in general not preserve the knot type.

In this section we want to consider the ensemble of unknotted SAP
(SAUK) weighted by their writhe. However, we will not use the same
approach as in \cite{universal_knots} to sample states via MCMC.
There are two reasons for this. First, when the writhe of SAPs is
weighted, the ensemble is dominated by increasingly denser states for
which the pivot algorithm becomes ineffective. Second, the ensemble
becomes increasingly populated by knotted states, thus it becomes hard
to sample effectively.

Instead, in this section we solve the model (\ref{eq:SAUK_Z-1}) via
simulations with the Wang-Landau algorithm (WLA) \cite{WANG_LANDAU_ORIG}
using a local move set that preserves the knot type. In this section,
we use $w=4\, Wr$, which is an integer for a SAP on the sc lattice.

\subsection{Algorithm and Data}

We use a parallel version of the WLA as discussed for example in \cite{PARADIGM_WL}.
We initialize an unknotted, rooted SAP on the positive half-space
of the sc lattice (without restriction to a slab geometry) and generate new states using the pull-moves \cite{hp-pull}
(excluding the end-move). The pull-moves certainly preserve the knot
type of the SAP, however in contrast to the case of the SAW, to our
knowledge, it is not proven that the pull moves are ergodic within
the knot type. For some simulations we also used kink transport between
random positions. Suppose the current state is $\varphi$, then we
obtain a state $\varphi^{*}$ by applying a move to $\varphi$. In
order to determine the writhe of $\varphi^{*}$, we rely on the formula~(\ref{eq:WR_sc_writhe_twist_formula}).
Therefore, we need to determine the linking number of the SAUK $\varphi$
with the pushed off unknot $\varphi+0.5\,\left(1,\,1,1\right)^{T}$.
The linking number can be determined by considering the signed crossing
in a projection. Suppose we project the SAUK into the $x_{2}$-$x_{3}$
plane, then all crossings will occur at potential crossing points
of the form $p_{c}=\left(x_{2}+0.5,\, x_{3}+0.5\right)$, $x_{2},\, x_{3}\in\mathbb{Z}$.
Therefore, every time a bond is added, the two potential crossing
points are determined and the bonds are linked to these crossing points
via pointers. When there are already bonds linked to the crossing
point, we cycle through all perpendicular bonds (that correspond to
the pushed off curve or vice versa) and compute the sum of signed
crossing that the new bond produces with all the bonds that have already
been linked. This corresponds to the change in linking number. The
same procedure is performed once a bond is removed. This time however,
the linking number changes by the negative of the sum of signed crossings
produced by the bond. This algorithm was inspired by \cite{LacherDataStructure}. 

In order to determine the extension (\ref{eq:WL_h-1}), we keep a
list of integer $N_{V}\left[H\right]$, which count the number of
vertices in the plane $x_{3}=H$. Then, every time a move removes
a vertex $\varphi_{i}$ from one position, we update the list as $N_{V}\left[\left(\varphi_{i}\right)_{3}\right]\leftarrow N_{V}\left[\left(\varphi_{i}\right)_{3}\right]-1$.
The list is modified accordingly when the vertex is placed onto its
new position. Finally, every time a list update yields $N_{V}\left[H\right]=0$
for some $H$, we set the new extension $h=H$. We do the same when
the population increases from $N_{V}\left[H\right]=0$. 

At given length $n$, we use the WLA to produce estimates for the
quantities 
\begin{equation}
s_{w\, h}:=\log C_{w\, h}\label{eq:swh}
\end{equation}
and
\begin{equation}
s_{w}\left(\beta_{h}\right):=\log\left(\sum_{h}C_{w\, h}\, e^{\beta_{h}h}\right).\label{eq:sw}
\end{equation}
Here, estimates means that we obtain (\ref{eq:swh}, \ref{eq:sw})
modulo a constant, including a statistical error and possibly a systematic
error. In order to obtain the estimate for $s_{w}\left(\beta_{h}\right)$
for general $\beta_{h}$, rather than for $\beta_{h}=0$ , we modify
the acceptance probability of the canonical WLA. Suppose the current
state at time $t$ is $\varphi\left(t\right)$ and the state $\varphi^{*}$
is proposed. Denote writhe and extension of the current state by $w\left(t\right)$
and $h\left(t\right)$, respectively. For $\varphi^{*}$, denote these
as $w^{*}$ and $h^{*}$, then, we accept $\varphi^{*}$ with the
probability
\begin{equation}
P_{acc}=exp\left\{ s_{w=w\left(t\right)}^{\left(est\right)}\left(\beta_{h}\, t\right)-s_{w=w^{*}}^{\left(est\right)}\left(\beta_{h},\, t\right)+\beta_{h}\cdot\left(h^{*}-h\left(t\right)\right)\right\} \label{eq:p_acc}.
\end{equation}

\begin{figure}[h]
\centering\includegraphics[width=10cm]{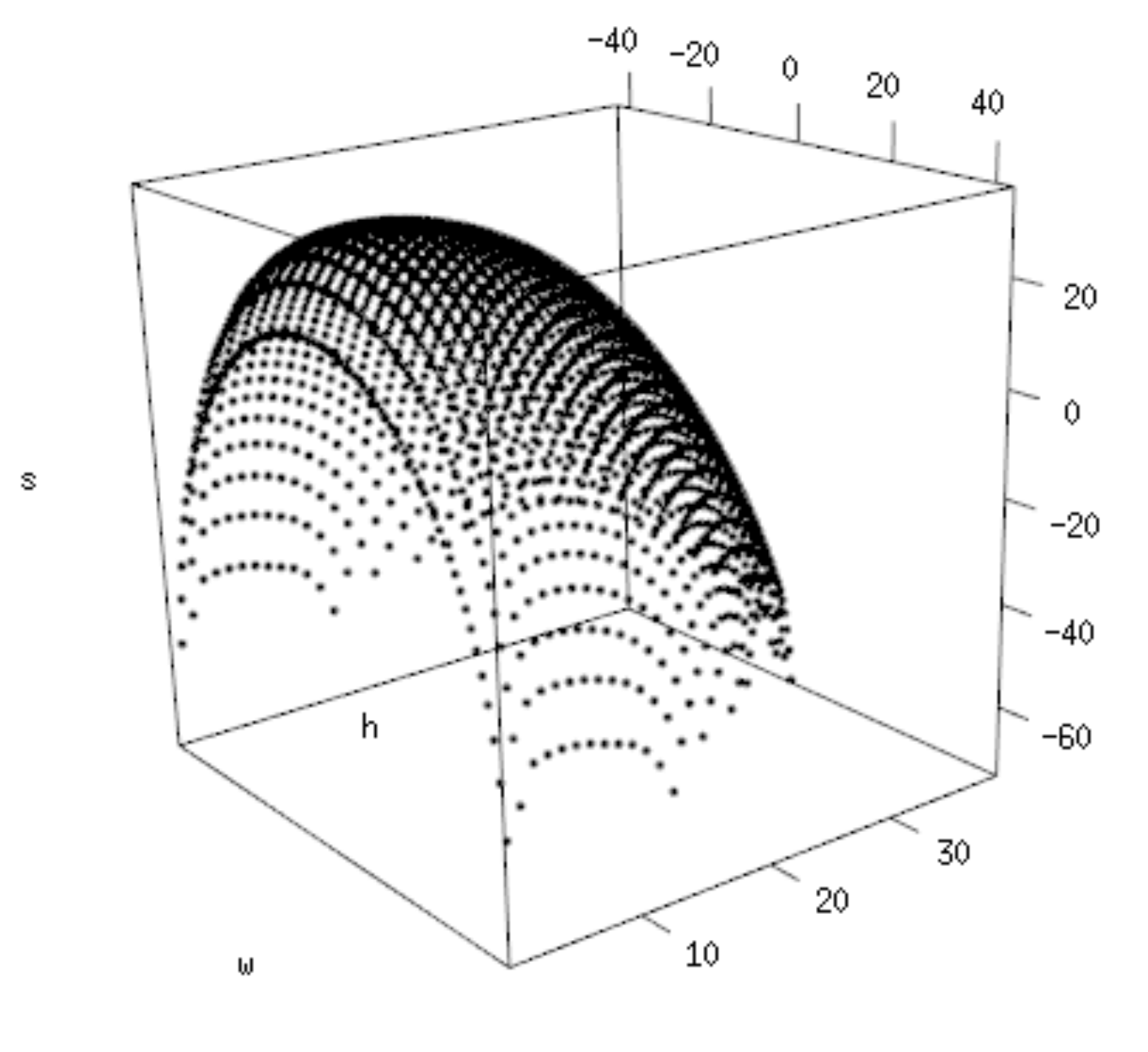}\caption{\label{fig:WL_2d_DOS-1}Two-dimensional entropy $s$ of the self-avoiding
unknot at $n=80$ over the extension $h$ and the writhe $w=4\, Wr$.
This entropy lacks macrostates $w>42$ as well as the macro state
$\left(w=0,\, h=0\right)$. The problem of generating samples of planar
polygons in three dimensions with the pull moves has been addressed
in the appendix of \cite{2258711620120301}.}
\end{figure}

We ran multiple simulations to estimate $s_{n\, w}\left(\beta_{h}\right)$
at different lengths $n$ and stretching forces $\beta_{h}$. We actually
produced estimates for $s_{n\,\left|w\right|}\left(\beta_{h}\right)$
and used the symmetry $s_{w}=s_{-w}$ to produce $s_{w}\left(\beta_{h}\right)$.
We usually used a cut-off for the writhe, so that $\left|w\right|\leq0.5\, n$.
This can be justified because $s_{n\, w}\left(\beta_{h}\right)$ falls
off quickly towards large $\left|w\right|$. This is for example shown
in the estimate of the two-dimensional entropy at $n=80$ in Figure~\ref{fig:WL_2d_DOS-1}.

The observables that we are interested in here are the derivatives
of the free energy with respect to $\beta_{l}$. We denote the estimates
of the derivatives by $f_{n}^{l}$, $f_{n}^{l\, l}$, $f_{n}^{l\, l\, l}$.
Therefore, 
\begin{eqnarray}
f_{n}^{l}: & = & n^{-1}\left\langle w\right\rangle _{n}^{\left(est\right)},\\
f_{n}^{l\, l}: & = & n^{-1}\left\langle \left(w-\left\langle w\right\rangle ^{\left(est\right)}\right)^{2}\right\rangle ^{\left(est\right)},\\
f_{n}^{l\, l\, l} & = & n^{-1}\left(\left\langle w^{3}\right\rangle ^{\left(est\right)}-3\left\langle w^{2}\right\rangle ^{\left(est\right)}\left\langle w\right\rangle ^{\left(est\right)}+2\,\left(\left\langle w\right\rangle ^{\left(est\right)}\right)^{3}\right),
\end{eqnarray}
where for example $\left\langle w\right\rangle _{n}^{\left(est\right)}\left(\beta_{l},\,\beta_{h}=const\right)$
can be computed as 
\begin{equation}
\left\langle w\right\rangle _{n}^{\left(est\right)}\left(\beta_{l},\,\beta_{h}=const\right)=\frac{\sum_{w}w\, e^{\beta_{l}\, w+s_{n\, w}^{\left(est\right)}\left(\beta_{h}\right)}}{\sum_{w}e^{\beta_{l}\, w+s_{n\, w}^{\left(est\right)}\left(\beta_{h}\right)}}.
\end{equation}

We provide a $95\%$ confidence interval for our results, by obtaining
from each simulation several estimates $s_{w}$ at different times.
We choose the times long enough apart so that the estimates can be
considered decorrelated. From each estimate, we compute observables,
pretend them to be independent and compute the standard confidence
interval. 

We note while it is possible to consider other observables like the
radius of gyration or the number of contacts by taking sample averages
along $\varphi\left(t\right)$, we will not present such results here.
We note that in particular good estimates for the radius of gyration
will be good only for short lengths as the local move set requires a
long time to decorrelate these observables. Also, since $\varphi\left(t\right)$
is not a Markov process, there is no (canonical) theory of the related
error. 

\subsection{Results}

\begin{figure}[h]
\centering\includegraphics[width=5cm]{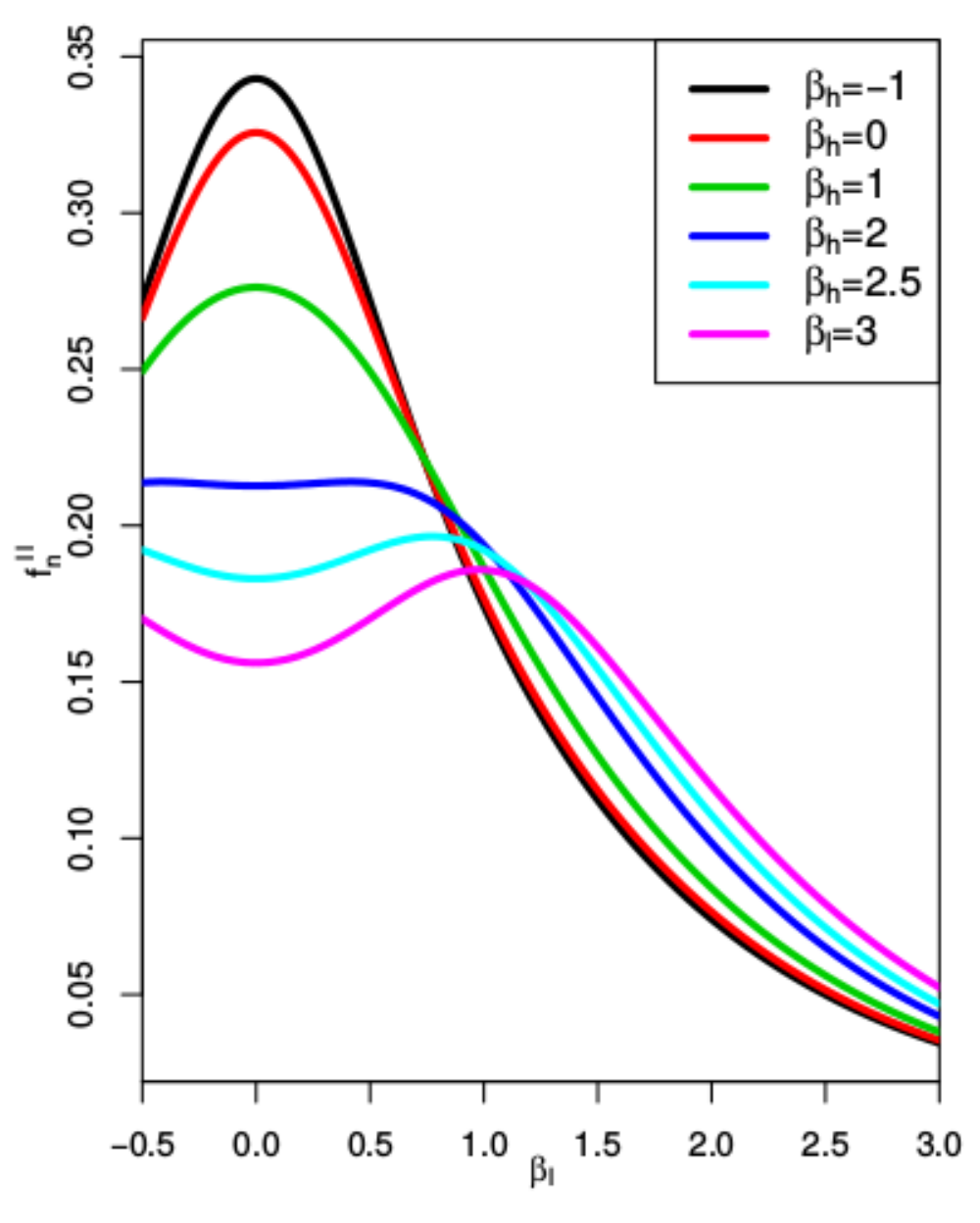}\caption{\label{fig:WL_2d_DOS-1-2}The graphic on the left-hand side shows
a heat plot (rainbow colors) of the writhe fluctuations in the $\beta_{l}-\beta_{h}$-plane
at length $n=80$. The fluctuations are maximal at the origin (violet-red).
As $\beta_{h}$ is increased, this maximum begins to branch. This
is also shown in the graphic on the right-hand side, which shows the
writhe fluctuations per length at selected pulling forces $\beta_{h}$
against the torque $\beta_{l}$.}
\end{figure}

Figure~\ref{fig:WL_2d_DOS-1-2} shows writhe fluctuations
$f_{n}^{ll}\left(\beta_{l},\,\beta_{h}\right)$ at length $n=80$.
At $\beta_{h}\leq0$, the maximum of the writhe fluctuation lies at
$\beta_{l}=0$. As the pulling force is increased, the maximum splits
into two modes, where the region between the peaks is expected to
be dominated by states with many bonds aligned in force direction.
This however suppresses the writhe fluctuations that occur along the
polymer.

At a pulling force of approximately $\beta_{h}=2.5$, two peaks
in $f_{n=80}^{ll}$ (Figure~\ref{fig:WL_2d_DOS-1-2}) appear. Note that
this pulling force is small enough so that we do not have to worry about lattice effects from fully extended walks. In fact, the
typical extension at $\beta_{l}=0,\,\beta_{h}=2.5$ is $\left\langle h\right\rangle _{n}/n\approx0.25$
and thus about half of the possible maximum of $\left(h_{n}/n\right)^{\left(max\right)}=\left(n-2\right)/\left(2n\right)$.
We will focus our analysis on $\beta_{h}=2.5$ which is an intermediate
pulling force.

\begin{figure}[h]
\centering\includegraphics[width=10cm]{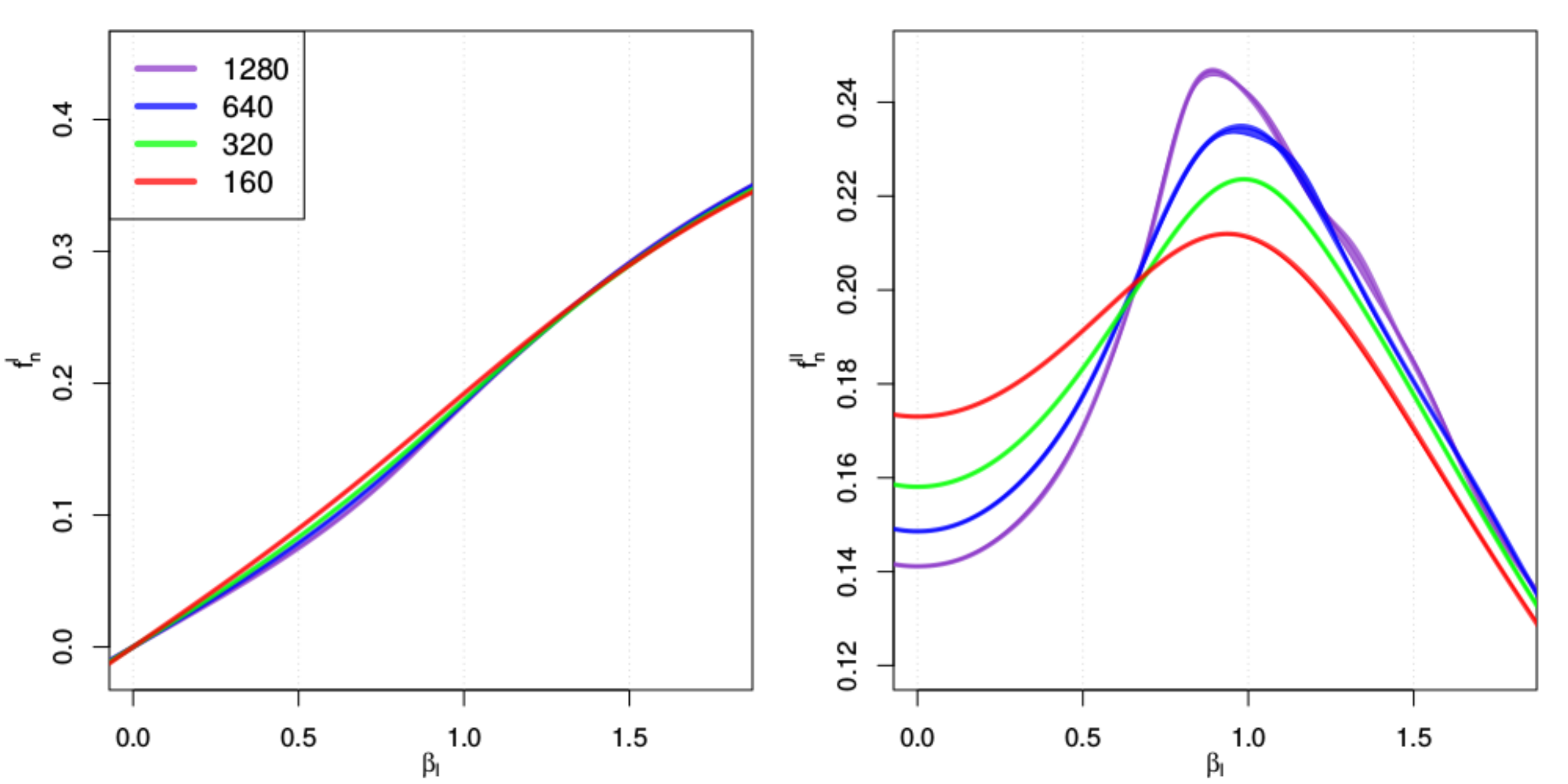}\caption{\label{fig:PULED_SAUK_SCALING}Scaling of the estimates of the first
and second derivative of the free energy $f_{n}$ with respect to
$\beta_{l}$ against $\beta_{l}$ at $\beta_{h}=2.5$.}
\end{figure}

Figure~\ref{fig:PULED_SAUK_SCALING} shows the scaling of the
first and second derivative of the finite size free energy with respect
to $\beta_{l}$ at $\beta_{h}=2.5$. While the scaling in the first
derivative is rather weak, the peak in the second derivative grows
with the length, however it is not clear whether it diverges.
To investigate the nature of this transition further, we consider the
third derivative.

\begin{figure}[h]
\centering\includegraphics[width=12cm]{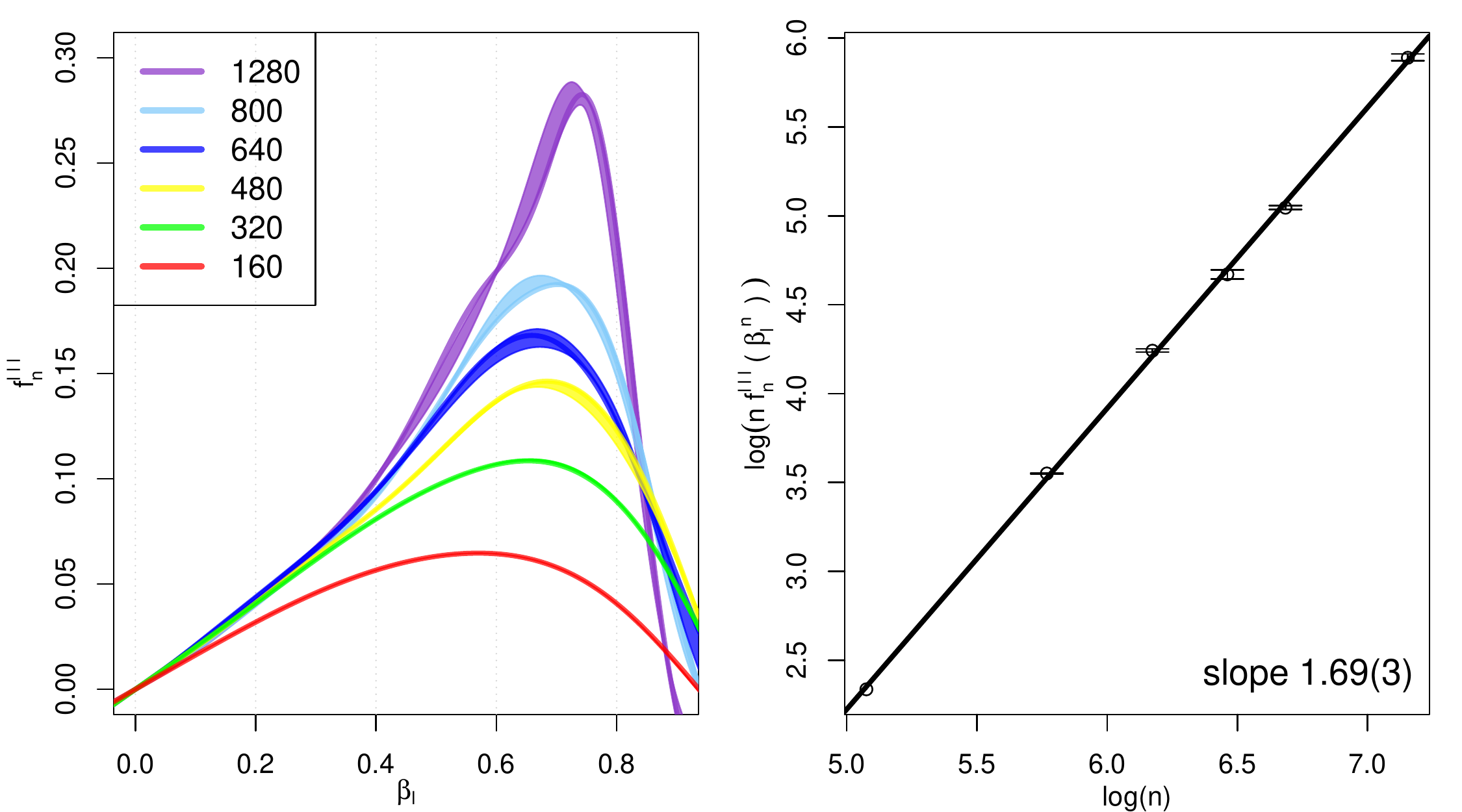}\caption{\label{fig:PULED_SAUK_SCALING_THIRD}Left graph: Scaling of the third
derivative of the free energy with respect to $\beta_{l}$ at $\beta_{h}=2.5$.
Right graph: Values of the maximum in $n\, f_{n}^{lll}\left(\beta_{h}=2.5\right)$
against the length in log-log-scale, together with a linear fit indicating a slope of $1.69(3)$. }
\end{figure}

The third derivative is shown in the first graph of Figure~\ref{fig:PULED_SAUK_SCALING_THIRD}.
The second graph in Figure~\ref{fig:PULED_SAUK_SCALING_THIRD} shows
a linear fit to the logarithm of the peak height in $n\, f_{n}^{lll}$
against the logarithm of the length. 
Using the standard scaling Ansatz~\ref{eq:INTRO_scaling_form} and differentiating three times with respect to $\tau$, we find that the third derivative, evaluated at the peak position, diverges as
$f_{n}^{lll}\left(\beta_{l}^{\left(n\right)}\right)\sim n^{3\phi-1}$. 

The good linear fit in Figure~\ref{fig:PULED_SAUK_SCALING_THIRD} suggests that this Ansatz is correct and we find
$3\phi-1=0.69(3)$, and hence $\phi=0.56(2)$. While this indicates that the crossover exponent might be slightly larger than $1/2$, we have not taken corrections to scaling
into account and hence don't feel confident enough to exclude the value $\phi=1/2$.

\subsection{Phase Diagram}

\begin{figure}[h]
\centering\includegraphics[width=8cm]{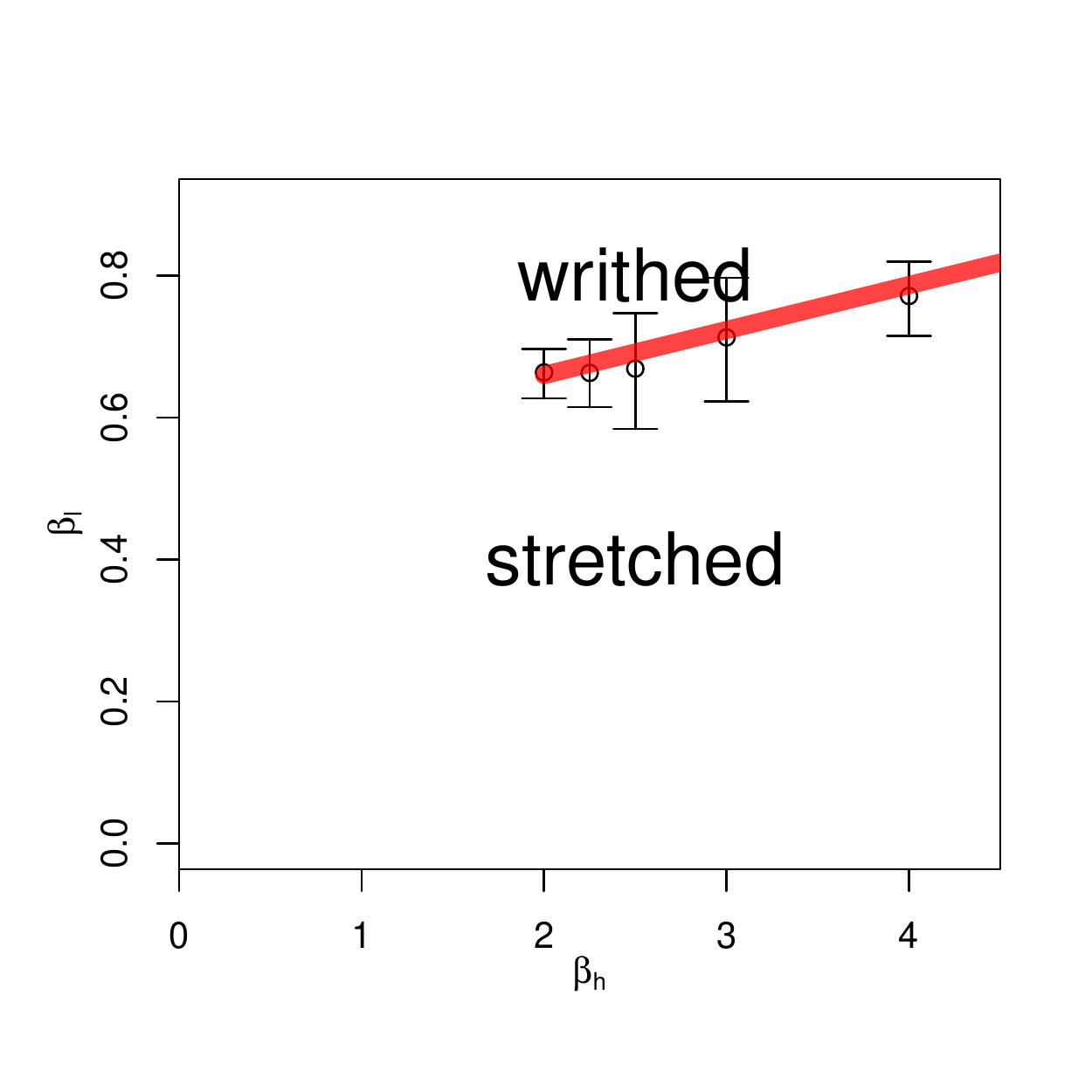}\caption{\label{fig:SAUK_PHASE_DIAGRAM}Phase diagram in the force-torque plane.
The red line marks a second-order phase transition. We cannot properly
determine the position of the transition for small pulling forces. }
\end{figure}

We use the peak position in the third derivative as a proxy for
the location of the finite-size transition. We use $n=640$ as reference
length and determine the location of the conjectured finite size transition
to lie at $\beta_{l}^{640}\left(\beta_{h}=2.5\right)=0.67\left(8\right).$
The error corresponds to the torque $\beta_{l}$ range for which the
confidence interval contains points that lie above the peak of its
lower boundary. Also note that for $n\geq480$, we do not observe
any shift of the peak location $\beta_{l}^{\left(n\right)}\left(\beta_{h}=2.5\right)$,
that lies outside the error. 
We obtained the locations of the transitions
at different $\beta_{h}$ to obtain the phase diagram in Figure~\ref{fig:SAUK_PHASE_DIAGRAM},
where we have called the phase at high torque 'writhed'. 

\begin{figure}[h]
\centering\includegraphics[width=12cm]{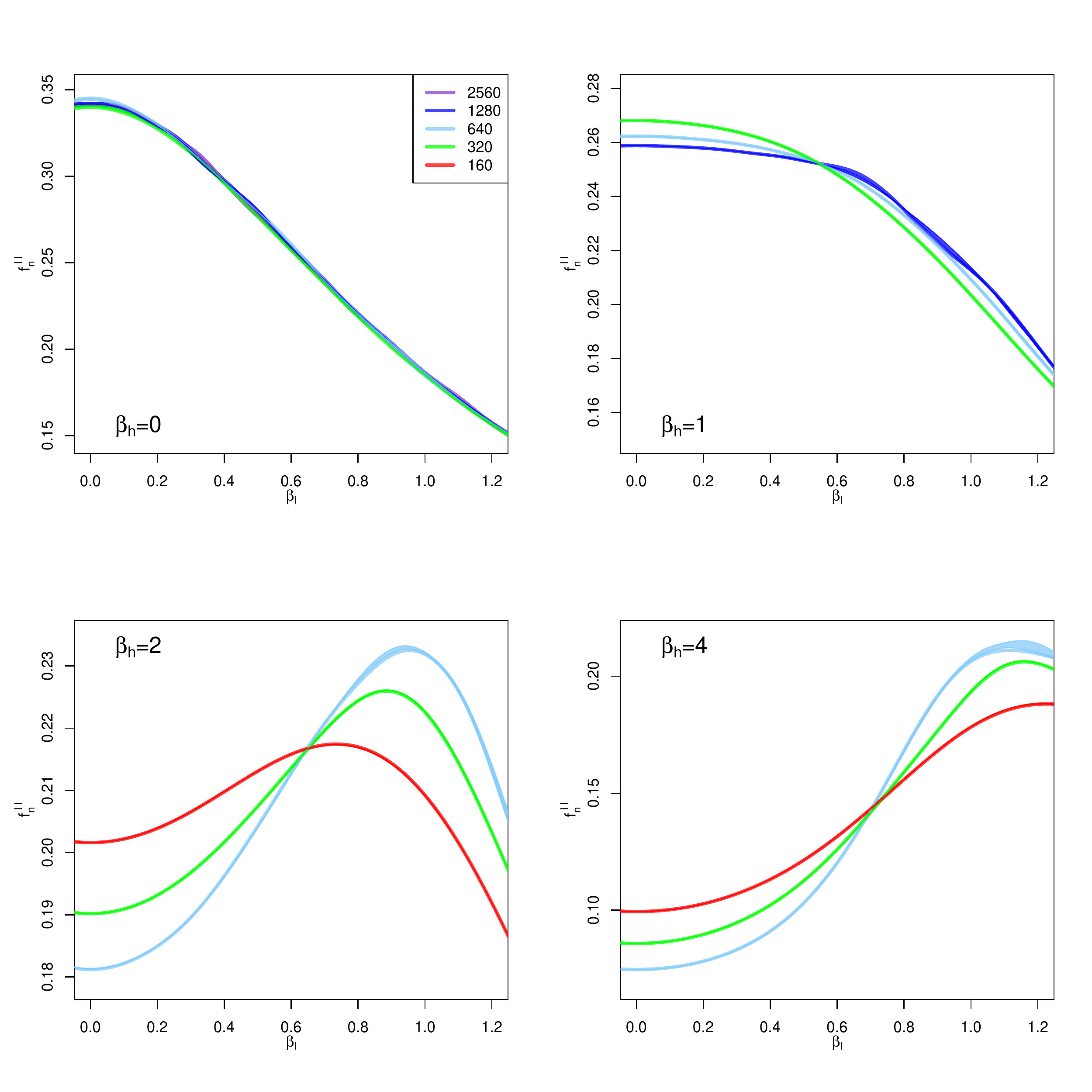}\caption{\label{fig:SCALING_fll_at_different_bh} Scaling of the second derivative
of the free energy with respect to $\beta_{l}$ at different pulling
forces. At $\beta_{h}=0$, the graph shows the curves for the length
$n=2580,\,1280,\,640,\,320$. One cannot make out clear scaling behavior.
At $\beta_{h}=1$ the considered lengths are $n=1280,\,640,\,320$.
There is weak non-trivial scaling so that the curve lowest at $\beta_{l}=0$
corresponds to $n=1280$ and the upper curve to $n=320$. At $\beta_{h}=2,\,4$
the lengths are $n=640,\,320,\,160$ and non-trivial scaling
is apparent.}
\end{figure}

At values $\beta_{h}<2$ we were unable to obtain good estimates of
the third derivative. Also, as $\beta_{h}$ decreases all signs of
scaling fade out from the second derivative at the considered lengths,
so that at $\beta_{h}=0$ the second derivative $f_{n}^{ll}$ appears
to scale trivially. \textbf{} This can be seen in Figure~\ref{fig:SCALING_fll_at_different_bh}
which shows the scaling of the second derivative at different pulling
forces.

Returning to $\beta_h=0$ we report that we have considered lengths of up to $n=2560$ and found no scaling.
We even probed (not well converged) length $n=5120$ and did not
see any good indication that non-trivial scaling might start to show. It is likely that then there is no transition when $\beta_h=0$, after all, the pulling force needs to be positive to have  stretched phase at high temperatures.

Hence, two scenarios present themselves. Firstly, the stretched to writhed transition exists for all non-zero positive values of $\beta_h$ and strong corrections to scaling inhibit our ability to detect the transition, or, secondly there exists  a minimum value $\beta_h$ below which the transition does not exist. 

Given the form of the writhe fluctuations being maximal around zero it is entirely possible that at small pulling force and short lengths the writhe fluctuations are dominated by a contribution that scales trivially
and overpowers the leading non-trivial scaling term. We therefore conclude the first scenario is more likely. However, further work on this is clearly warranted.

\section{Conclusion}

In this paper we study an ensemble of unknotted self-avoiding polygons on the sc lattice weighted by writhe
to identify potential phase transitions, as this can be related to
experiments that turn DNA. We define and solve a simplified model for the strongly pulled regime, and show in this model the
existence of a phase transition between phases of low and high writhe at a non-zero temperature. 

We next consider the full problem using simulations with
the Wang-Landau Algorithm. When the unknotted self-avoiding polygon is pulled sufficiently, we provide results
that are compatible with a second-order phase transition between phases of low and high writhe.
At low torques the unknotted self-avoiding polygon is in a stretched phase with anisotropic size scaling. As the pulling force is reduced, 
we observe that the scaling in the writhe fluctuations fades out, so that at
zero pulling force $\beta_{h}=0$,  we do not find any signs of scaling. 
It will be interesting to explore in further studies how the transition changes at low pulling forces.

\section*{Acknowledgments}

One of the authors, ED, gratefully acknowledges the financial support
of the University of Melbourne via its Melbourne International Research
Scholarships scheme. Financial support from the Australian Research
Council via its support for the Centre of Excellence for Mathematics
and Statistics of Complex Systems and the Discovery Projects scheme (DP160103562)
is gratefully acknowledged by one of the authors, ALO, who also 
thanks the School of Mathematical Sciences, Queen Mary University of London 
for hospitality. ED also acknowledges support from VLSCI HPC and Edward HPC for providing computational resources. 

\section*{References}


\end{document}